\documentclass[]{raa}
\usepackage[utf8]{inputenc}
\usepackage{amsmath,esint}
\usepackage{amsfonts}
\usepackage{amssymb}
\usepackage{graphicx}
\usepackage{mathrsfs}
\usepackage{framed}

\usepackage{natbib}
\bibpunct{(}{)}{;}{a}{}{,} % to follow the A&A style

\usepackage[]{hyperref}
\hypersetup{pdftitle = The title of my PDF, pdfauthor = My name, pdfsubject= The subject, pdfkeywords = keyword1 keyword2 keyword3} 
\hypersetup{colorlinks = true, linkcolor = green, anchorcolor = red, citecolor = blue, filecolor = red, pagecolor = red, urlcolor = red}

\begin{document}

\title{Quark Nova Model for Fast Radio Bursts
}
%   \subtitle{I. Place Your Subtitle Here}

   \volnopage{Vol.0 (200x) No.0, 000--000}      %%preserved for Editor. DOn't remove!
   \setcounter{page}{1}          %%starting page, preserved for Editor. DOn't remove!

   \author{Zachary Shand
      \inst{}
   \and Amir Ouyed
      \inst{}
   \and Nico Koning
      \inst{}
   \and Rachid Ouyed
      \inst{}
	}
   \institute{Department of Physics and Astronomy, University of Calgary, 2500 University Drive NW, Calgary AB T2N 1N4, Canada {\it rouyed@ucalgary.ca}\\
   }

   \date{Received~~2009 month day; accepted~~2009~~month day}

\abstract{
FRBs are puzzling, millisecond, energetic radio  
transients with no discernible source; observations 
show no counterparts 
in other frequency bands. 
The birth of a quark star from a parent neutron star 
experiencing a quark nova
- previously thought
undetectable when born in isolation - provides a natural 
explanation for the emission characteristics of FRBs. The 
generation of unstable r-process elements in the 
quark nova ejecta provides millisecond 
exponential injection of electrons into the surrounding 
strong magnetic field at the parent neutron star's light cylinder via $\beta$-decay. 
This radio synchrotron emission has a total duration of hundreds of milliseconds 
and matches the observed spectrum while reducing the inferred dispersion measure 
by approximately 200 cm$^{-3}$ pc. The model allows indirect measurement 
of neutron star magnetic fields and periods in addition to providing 
astronomical measurements of $\beta$-decay chains of unstable neutron rich nuclei. 
Using this model, we can calculate expected FRB average energies ($\sim$ 10$^{41}$~ergs), spectra shapes and provide a theoretical framework for determining 
distances. 
}

   \authorrunning{Shand et al. }   %author_head in even pages
   \titlerunning{QNe Model for FRBs }  % title_head in odd pages

\maketitle

\section{Introduction}
\label{sec:intro}
The observation of fast radio bursts (FRBs) has created an 
interesting puzzle for astronomers. Explaining the origin of these
millisecond bursts has proven difficult because no 
progenitor has been discovered and there is no detectable emission 
in other wavelengths. 
Many possible explanations for these bursts have been put forward: 
blitzars \citep{blitzars}, 
neutron star mergers \citep{NSMMergerFRB},
magnetars \citep{Magnetars2007,Magnetars2013,Magnetars2014,Magnetars2015}, 
asteroid collisions with neutron stars \citep{AsteroidFRB}, and
dark matter induced collapse of neutron stars \citep{DarkMatterCollapse}, to name 
a few; however,
none so far have been able to explain the many puzzling features of the 
emission: the signal dispersion, energetics, polarization, galactic rate,
emission frequency, and lack of (visible) companion/source object. 

Direct observational evidence for quark stars is difficult to accrue due to some of their
similarities to neutron stars (e.g. compactness, high magnetic field, low luminosity). In our
model, FRBs signal the birth of a quark star and as more FRB observations become available,
they may be able to provide evidence for the existence of quark stars. Other theoretical
observational signatures of quark stars include: 
gravitational waves from quark star oscillation modes \citep{2003A&A...412..777G,0264-9381-31-15-155002}; 
gravitational waves from a strange-quark planet and quark star binary system \citep{2015ApJ...804...21G}; 
equation of state constraints of the mass-radius relationship \citep{2015arXiv150902131D} 
and high-energy emission directly from the quark star in X-ray and gamma ray emission \citep{0004-637X-729-1-60}.

The Quark Nova (hereafter QN) model has many 
ingredients which prove useful when trying to explain FRBs. Firstly, the parent 
neutron star has a strong magnetic field which 
can generate radio synchrotron emission. Secondly, the 
neutron rich ejecta of the QN has a unique mechanism for producing 
these electrons via $\beta$-decay 
of nuclei following r-process nucleosynthesis in the QN ejecta \citep{Jaikumar2007,rJava1,rJava2}. 
Our model makes use of $\beta$-decay of nuclei in the ejecta of a QN to 
produce radio synchrotron emission. The exponential decay 
inherent to the nuclear decay explains the millisecond pulses observed 
in a way that no other model so far can. 

As an isolated, old neutron star spins down the central 
density increases. Neutron star spin-down, which increases the core density leading to quark deconfinement, in conjunction with 
accretion and s-quark seeding can 
trigger an explosive phase transition \citep{Ouyed2002,Staff2006,OuyedCombustion}. 
This explosion ejects a dense layer of neutrons which produce unstable 
heavy nuclei \citep{Keranen2005Ejecta,Ouyed2009Evolution}. These unstable nuclei decay and become observable at 
the light cylinder (hereafter LC) where they emit 
synchrotron emission due to the strong magnetic field.
(For a 10~s period and $10^{12}$~G parent neutron star this corresponds to $B \sim 300 \mbox{~G}$ at the LC.) This generates a 
unique geometry and field configuration where our 
emission occurs at the intersection of a cylinder and sphere 
in space traced out by the ejecta as it  passes through 
the LC. 

We expect FRBs to be associated with old, isolated neutron stars and not 
young, rapidly rotating neutron stars. Neutron stars born with short periods 
quickly spin down and increase their central densities and, when these neutron stars are
born with sufficient mass, deconfinement densities are reached very quickly (within days to weeks).
Instead of emitting as an FRB, these QN will interact with the supernova ejecta leading to either a 
super-luminous supernova \citep{SLSNKostka2014}, or a double-humped 
supernova \citep{1674-4527-13-12-007}. This
interaction with the surrounding material is expected to either disrupt the FRB emission, or bury 
the signal in the surrounding dense ejecta. The remaining neutron stars which are born massive  
and with high periods will take much longer (millions of years) to achieve the required critical 
core density. As such, 
we expect a continuum of FRB progenitors with a range of periods and magnetic fields to 
be emitting at various frequencies and luminosities. 

Two important consequences of this scenario are: 1) an extended time spectra as 
the sphere sweeps across higher and higher latitudes of the LC and
2) an exponential time profile of millisecond duration from decay of $\beta$-decaying nuclei. As is demonstrated later (in 
Fig.~\ref{fig:sourceSpectrograph}), these qualities match observations of FRBs and 
allow us to provide testable predictions for distances, energies, rates and 
bandwidth of future FRB measurements which we can link to 
old, isolated neutron star populations and nuclear rates.

\section{Model Overview}
As a neutron star ages, spin-down increases its central density. 
For massive enough 
neutron stars, this can cause a phase transition birthing a quark star.
The exploding neutron star ejects - at relativistic speeds - the outer layers of the 
neutron rich crust. The ejecta is converted to unstable r-process material 
in a fraction of a 
second and reaches a stationary r-process distribution which is maintained by 
the high-density of the ejecta and the latent emission of neutrons 
from fission (fission recycling). 
This distribution of unstable nuclei, which normally would decay within seconds, 
can continue to exist while the density is high and 
fissionable elements are able to release 
neutrons back into the system. This stationary distribution will be disrupted 
by decompression (or other disruption) of the QN ejecta as it passes 
through the LC of the surrounding parent neutron star magnetic field. 

Once the ejecta is beyond the current sheets of the LC, the QN ejecta 
will no longer support a stationary r-process distribution and electromagnetic radiation 
from the ejecta will no longer be attenuated by the LC 
plasma, becoming visible \footnote{The current sheets at the 
light cylinder produce a Faraday cage effect which 
shields the emission of $\beta$-decay occurring before the 
r-process material crosses the LC.}.
The nuclei will decay via a rapid series of $\beta$-decays. These generated electrons are emitted with MeV energies in a strong magnetic field 
producing radio synchrotron emission. The frequency and power of this emission 
depends, of course, on the magnetic field strength at the LC where the QN ejecta 
crosses it. As a result, the emission spectrum is controlled by the geometry of the spherical intersection 
of the ejecta with the LC and allows inference of the progenitor neutron star's period 
and magnetic field.

\begin{figure}
	\centering
	\includegraphics[scale=1.0]{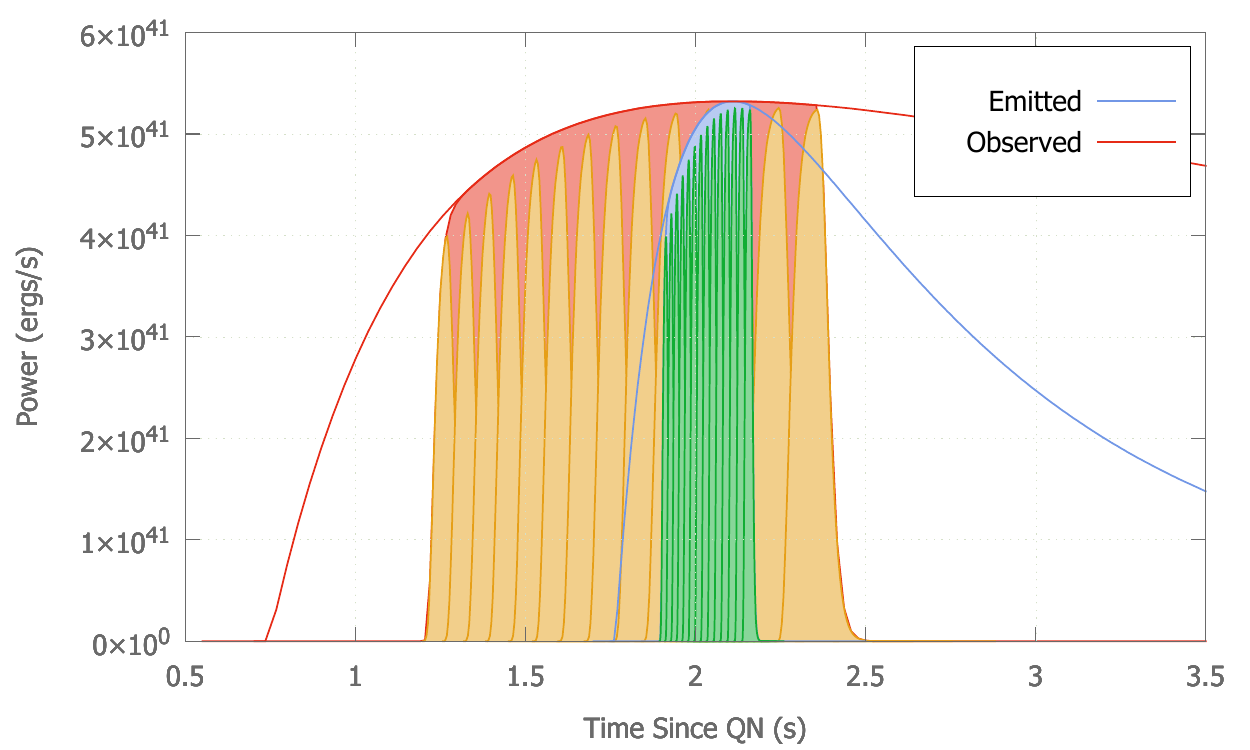}
	\caption{ Light curve of the FRB emission at the source (blue) and the observed emission with dispersion smearing (red). 
	The filled portion of the curves indicates the viewable emission for bandwidth limited observation in the range of 1175--1525 MHz.  
	The green and orange pulses are 25 MHz sub-bands of the signal which exhibit the characteristic millisecond pulse duration.
	\label{fig:lightCurve}}
\end{figure}

A distinguishing detail in our model is the duration of the emission, which has a duration 
of seconds instead of milliseconds. The signal is a continuous series of synchrotron emissions from bunches of electrons generated from $\beta$-decay,
which produce coherent, millisecond duration pulses at each frequency over several  
seconds (or hundreds of milliseconds for a bandwidth limited observation), as can be seen in Fig.~\ref{fig:lightCurve}.
This is in apparent contrast with observations, which 
indicate a total emission duration of milliseconds. This discrepancy in the FRB duration stems from the de-dispersion process 
which stacks all frequency channels to a common initial time (as is done in pulsar de-dispersion and signal folding). As such, according to our model, 
the FRB emits a series of coherent, self-similar signals at different frequencies which have been erroneously stacked as a time synchronized 
pulse: the total duration of the FRB is actually several seconds long and the observed signal duration is still
dominated by cold plasma dispersion, as is demonstrated by the red/orange signal in Fig.~\ref{fig:lightCurve}.

The model comprises a combination of relatively standard physics phenomena:
$\beta$-decay, magnetic field geometry and synchrotron emission. In composite, they 
generate a unique emission mechanism reminiscent of 
FRBs. The following sections (Sections \ref{sec:firstPiece} to \ref{sec:lastPiece}) will be spent 
outlining each component of the model in greater detail and setting out the 
relevant equations necessary to generate the equations which describe the 
synchrotron emission spectra of the r-process material generated in the QN ejecta 
(see Section \ref{sec:puttingItTogether}).

\subsection{R-process Elements}
\label{sec:firstPiece}
\begin{figure}
	\centering
	\includegraphics[width=\linewidth]{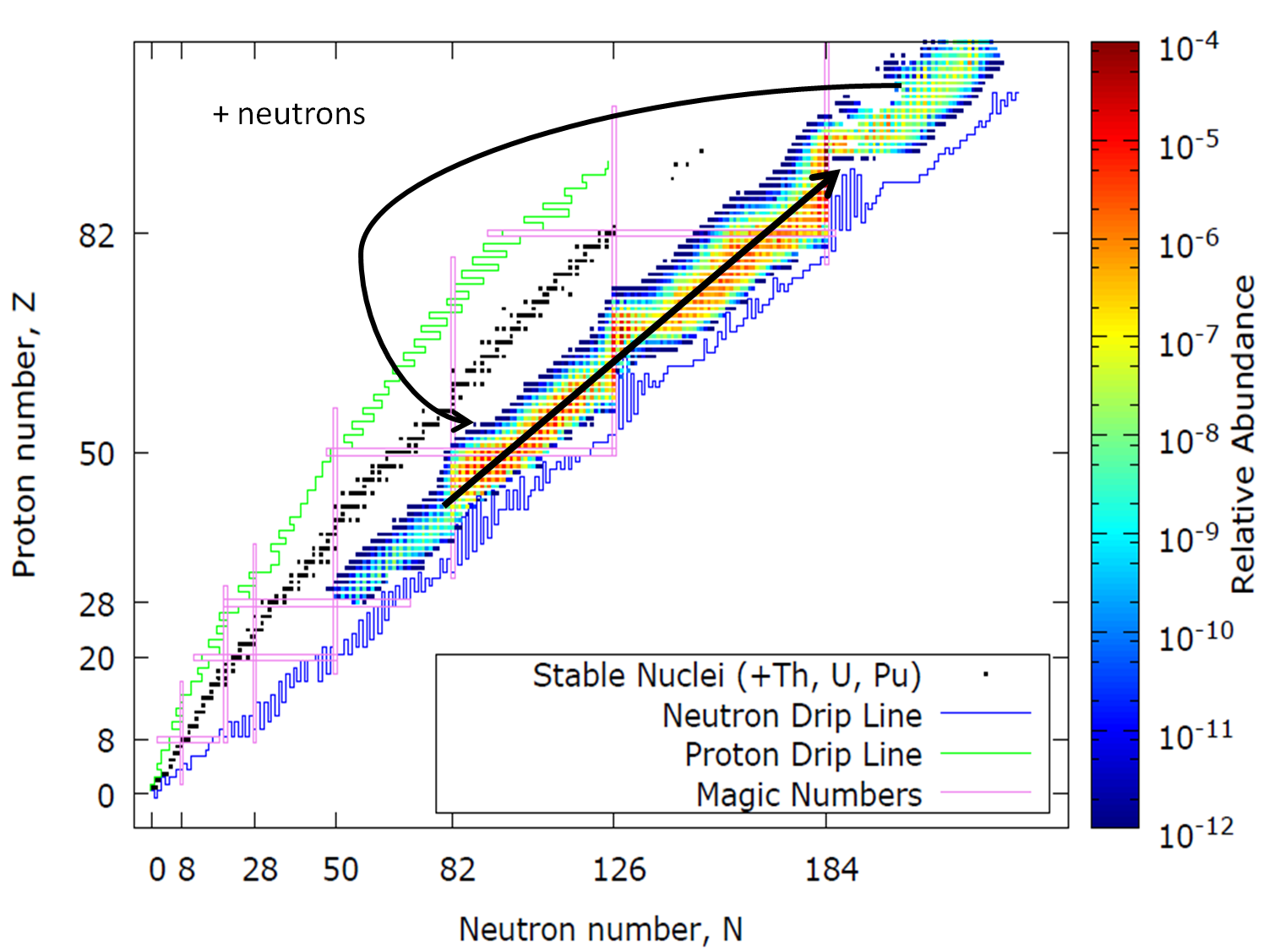}
	\caption{Stationary distribution of the r-process near the end 
	of the r-process. The neutron source becomes largely depleted 
	and no major changes to the distribution occur; however, 
	fissionable nuclei emit neutrons and create a cyclic recycling 
	process which can maintain a distribution of unstable nuclei 
	for several seconds ($>10$~s) in contrast to the 
	milliseconds required for generation of r-process material. 
	The distribution is computed using r-Java 2.0 as in 
	\citet{rJava2}. \label{fig:rDist}}
\end{figure}

The formation of heavy elements by the r-process is one of the 
predictions of the QN model \citep{Jaikumar2007,rJava1,rJava2}. Due to large number of 
free neutrons delivered by the ejected neutron star crust, it is possible 
for heavy fissile nuclei to be created during the r-process. 
R-process calculations show that superheavy ($Z>92$) fissile elements 
can be produced during the neutron captures. This leads to fission 
recycling during the r-process \citep{rJava2}. Each fission 
event ejects (or releases) several ($\sim 5-10$) neutrons back into the r-processing 
system. (This is due to neutron evaporation during asymmetric fission 
and is same mechanism present in 
nuclear reactors.)
 Because of this, once these fissionable elements are produced, we 
are guaranteed to have free neutrons continually resupplied to the system 
which recombine into heavier elements which will in turn fission again 
and create a feedback loop in the r-process inside the QN ejecta. 
This allows for continued neutron captures and ``r-processing'' much longer than 
traditionally expected (see Fig.~\ref{fig:rDist}). These unstable isotopes 
are then able to supply us with the electrons which will decay in coherent bunches over 
different magnetic fields to produce the final aggregate emission spectrum.

\subsubsection{R-process Rates}
\begin{figure}
\centering
\includegraphics[width=\linewidth]{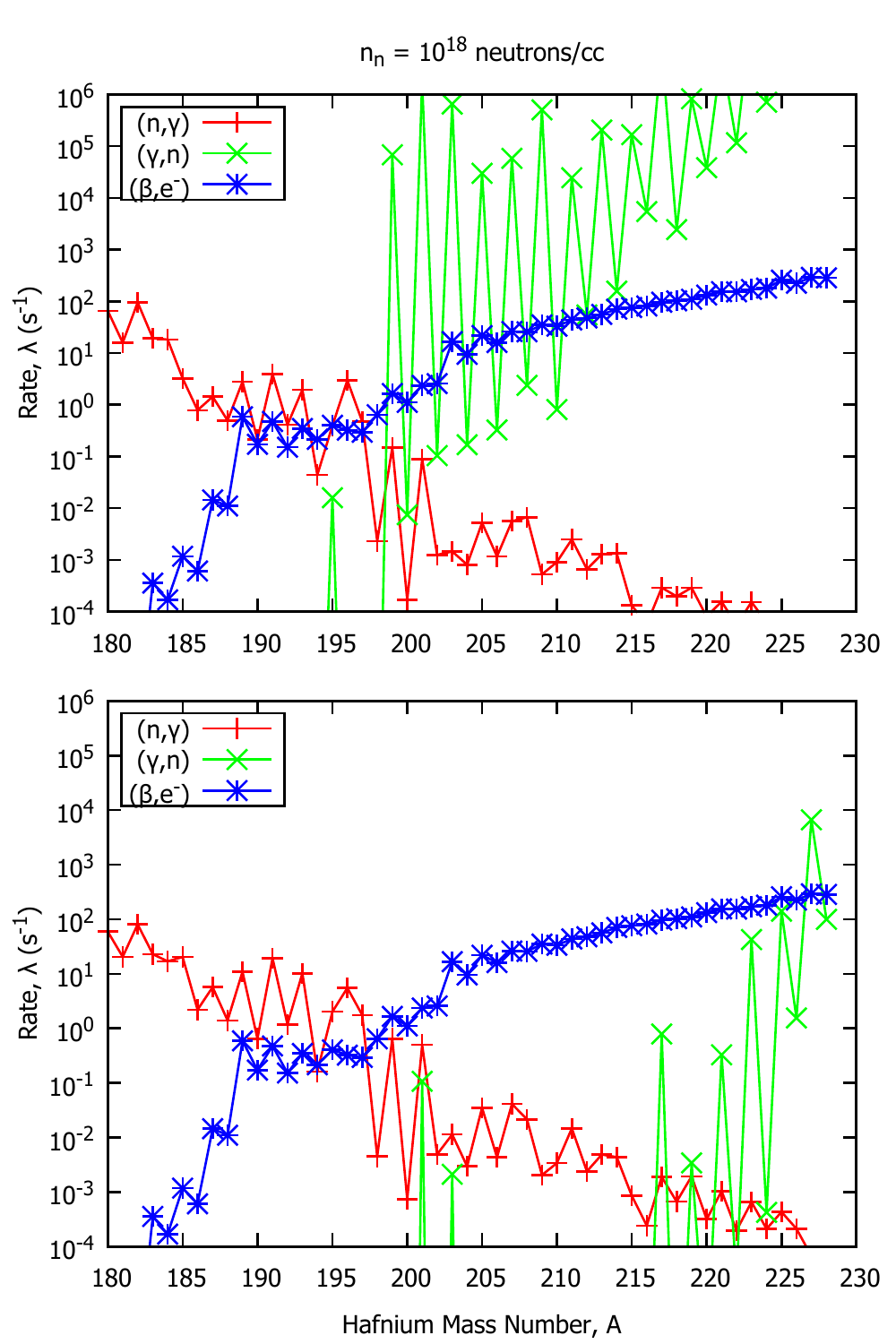}
\caption{Relevant temperature dependent rates 
for hafnium show different regimes in the r-process. The 
neutron density is $10^{18}$cm$^{-3}$ and the temperatures 
are $1.0\times 10^{9}$K (top) and $0.5 \times 10^9$K (bottom).  \label{fig:competeBeta}}
\end{figure}

The basic r-process mechanism operates as a competition between neutron capture 
reactions (n,~$\gamma$) and its inverse photo-dissociation ($\gamma$,~n) with the 
much slower $\beta$-decays slowly building up each successive isotope. This 
picture holds true for most of the r-process; however, as the gas 
expands and cools, the photo-dissociation rates fall off very quickly with 
temperature. The neutron-capture rates, on the other hand, do not have 
as strong a temperature dependence and are mainly a function of neutron density
(see Fig.~\ref{fig:competeBeta}).

For an initially dense ejecta (which is of course the case for a neutron 
star crust ($\rho \sim 10^{14}$~g~cm$^{-3}$) this means that neutron captures can continue 
at low temperatures even after significant initial decompression.
Instead of competing with photo-dissociations, 
they will begin to compete with $\beta$-decays instead as shown in 
Fig~\ref{fig:competeBeta}. At this point, 
simulations show that the isotopic distribution enters an approximate steady 
state (possible because of the fission recycling) which can 
continue until the density and subsequently the neutron capture rates lower
sufficiently. The rates shown here come from \citet{MollerBeta} and the BRUSLIB database \citep{HFBrates1,HFBrates2}. These rates are all theoretical due to 
the difficulty of measuring them in the lab \citep{Reiner2011} 
and presents a way of observationally constraining $\beta$-decay rates 
by measuring synchrotron radiation from theses decay chains. 

\subsubsection{Decay Chain of Unstable Isotopes}
\begin{figure}
\centering
\includegraphics[width=\linewidth]{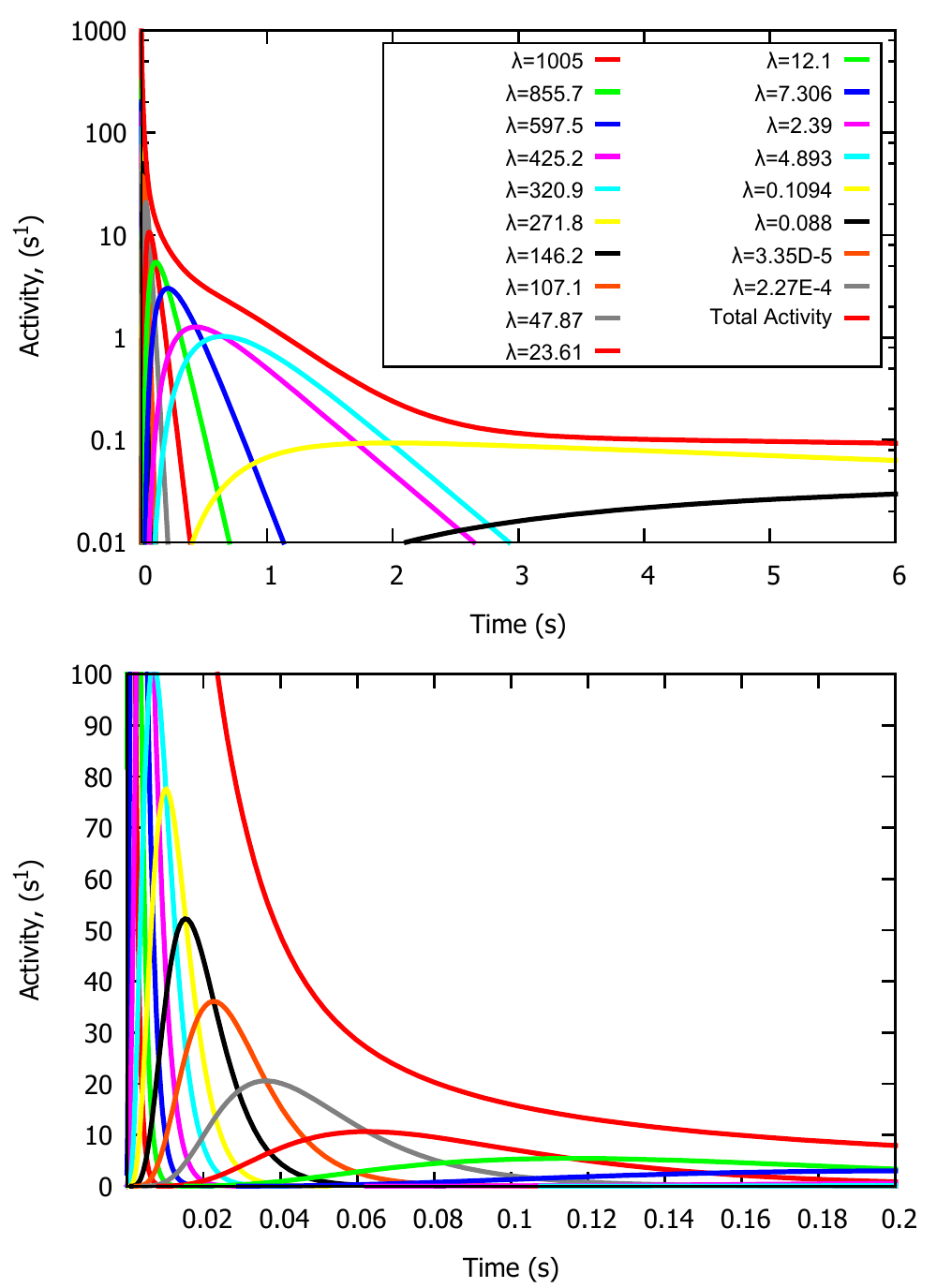}
\caption{Solution of the Bateman equation for the A=156 
$\beta$-decay chain beginning near the drip-line. Once 
the neutron decays stop competing with $\beta$-decays at 
the end of the r-process (when the density 
falls sufficiently), the activity of the material will 
$\beta$-decay with this characteristic exponential behaviour. 
The top panel is logarithmic in the activity (y-axis) 
and the bottom panel is the very early portion of the above panel without the 
logarithmic y-axis.\label{fig:betaChain}}
\end{figure}

For elements lighter than the actinides, $\beta$-decay (or electron capture) 
is the dominant decay mode available to them. These decay chains can
be described by coupled differential equations which have a simple analytic
solution known as the Bateman equations \citep{Krane}:
\begin{equation}
 \mathscr{A}_n = N_0 \sum_{i=1}^n c_i e^{-\lambda_it},
 \label{eq:nucDecay}
\end{equation}
\begin{equation}
c_m = \frac{\prod_{i=1}^n \lambda_i}{\prod_{i=1}^n \prime (\lambda_i-\lambda_m)},
\end{equation}
where the prime indicates that the lower product excludes the term where i=m.
This equation gives the activity (number of events per unit time) of a 
decay chain. For a purely $\beta$-decay chain this gives 
the rate at which free electrons are produced by nuclei. 
As can be seen in Fig.~\ref{fig:betaChain}, once $\beta$-decay dominates, 
the unstable nuclei will decay roughly exponentially and have 
a short burst of activity that will emit synchrotron radiation 
if subject to an external magnetic field. This naturally gives us 
an exponential fall-off time once neutron captures stop, which 
as discussed above, can be much later in the expansion than traditionally 
expected in the r-process due to the high density of the ejected material
(which is abruptly triggered by decompression across the LC).
Once emitted, the electrons will recombine with the surrounding weakly 
ionized plasma which should prompty emit in the keV range which could be 
visible if the plasma is diffuse enough; however, at much lower total 
emission power than the FRB burst itself (likely making it undetectable). 

\subsection{Geometry and Field Configuration}
\begin{figure}
	\centering
	\includegraphics[width=\linewidth]{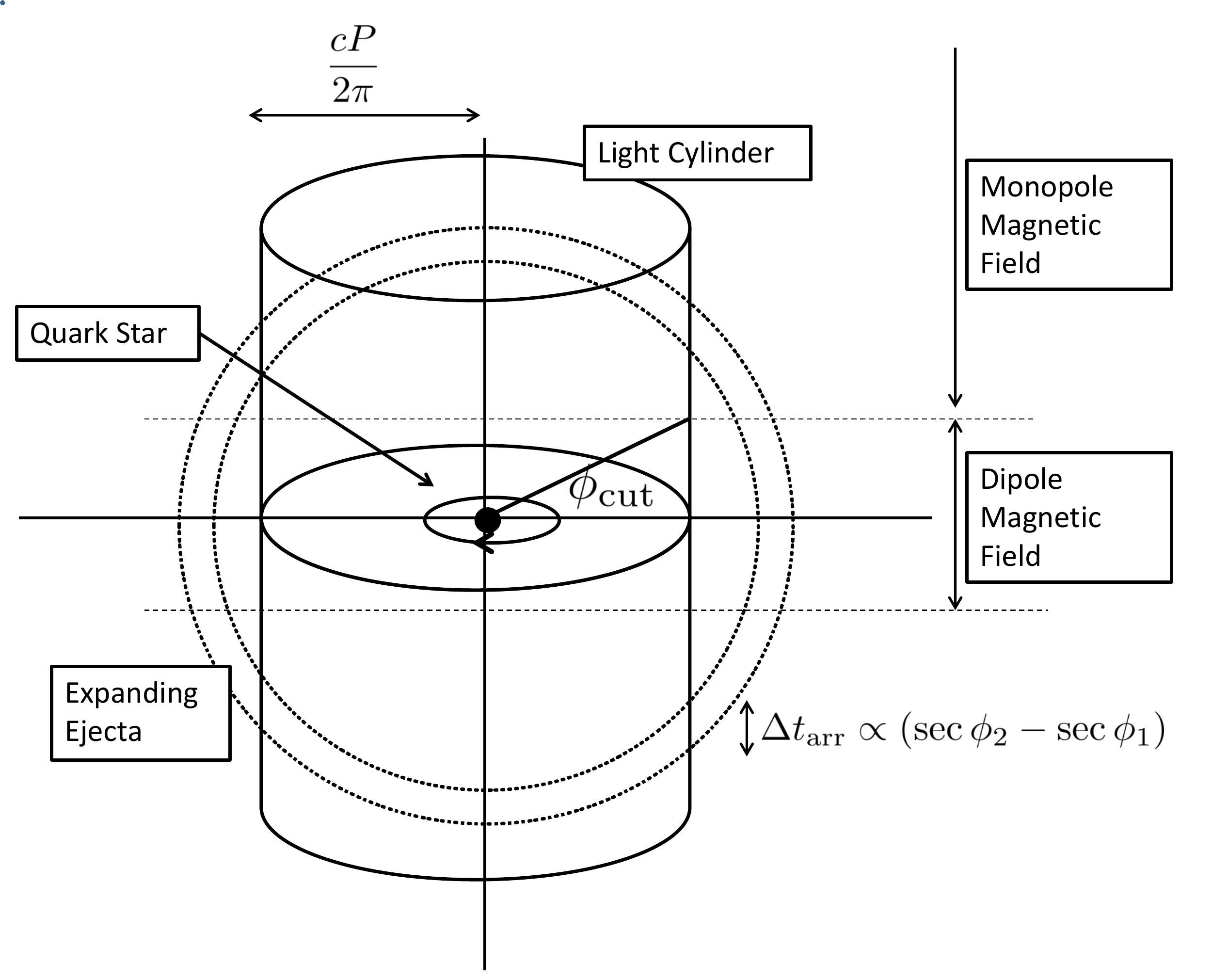}
	\caption{Diagram of outlining geometry of the model. The latitiude angle angle is 
	specified relative to the quark star location and depicts the angle at 
	which the spherical QN ejecta intersects with the LC.
	The figure depicts two intersections of the ejcta with the cylinder 
	at two different latitude angles separated by the arrival time 
	of the two latitue angles.  \label{fig:cylinderDiagram}}
\end{figure}

Assuming that the emission is triggered at the LC, we need to 
define a coordinate system with which to describe emission time and 
magnetic field strength (see Fig.~\ref{fig:cylinderDiagram}).
The LC is simply defined as the point where a co-rotating magnetic 
field would be rotating at the speed of light. The magnetic field configuration 
has a breakpoint in it where the magnetic field lines no longer reconnect and 
fall off as a mono-polar magnetic field. For a LC radius defined 
by: 
$r_{LC} = \frac{c P}{2 \pi}$,
where c is the speed of light and P is the period, the magnetic field strength at 
the LC at latitude, $\phi$, is given by: 
\begin{equation}
	B = B_0 \left( \frac{r_{0} \cos\phi }{r_{LC}} \right)^a ,
		\begin{cases}
		a = 3 & ~0^\circ \leq \phi < 30^\circ \\
		a = 2 & 30^\circ \leq \phi < 90^\circ. \\
		\end{cases}
\end{equation}
This will cause radiation in the equatorial plane, where the dipolar 
field configuration exists, to be a factor of $c$ weaker ($c^2$ for power). 

For a relativistic, spherical shell this also provides a straightforward description 
for the arrival time of the material at the LC: 
\begin{equation}
\label{eq:arrivalTime}
	t_\textrm{arr} = t - \frac{P}{2 \pi \beta} \sec \phi,
\end{equation}
which provides a time delay between emission at different latitude angles. 

\subsection{Light Cylinder Details and Magnetic Field}
The magnetic field configuration outside due to the LC 
will transition to a monopole magnetic field for an aligned rotator \citep{Michel1973,CFK1999} 
or will have more complicated shape for an unaligned rotator \citep{Spikovsky2006}. 
As the QN 
ejecta passes through this region, the magnetic field will be disrupted 
and the details of the dynamics would be complicated to simulate; however, 
the disruption of the currents and the change of the magnetic field 
are assumed to cause a disruption of the QN ejecta. The details of the expansion across 
the LC are beyond the scope of this paper; however, as the ejecta passes 
through the boundary, there will be decompression of the material as the magnetic field 
distribution changes and the currents at the LC will interact with the 
matter and be disrupted. Since the LC inhibits emission of electromagnetic radiation across the LC \citep{LCProperties}, 
this physical disruption can help to explain the sudden rise of the emission of the 
FRBs. 

Both the neutron star and quark star will have a LC and so there 
is some possible ambiguity in what magnetic field will be present at the LC when the ejecta crosses it. Since QN ejecta is highly relativistic ($v\sim 0.99c$) we expect it to reach the LC of the parent neutron 
star before there is time for the quark star to fully rearrange the magnetic field around it which occurs much 
slower \citep{OuyedFieldRalignment}.
This can allow inference of properties of the neutron star population which 
undergo quark-nova and emit these FRBs. 

\subsection{Synchrotron Emission}
\label{sec:lastPiece}
The emitted electrons are assumed to have a constant $\gamma$ across all decaying 
elements. This is done both to simplify the model and also due to a lack of 
nuclear data of $\beta$-decay electron energy distributions. If we look at the 
flux density, this factor is suppressed and so the assumption ends up not 
factoring into comparisons for flux density (see Fig.~\ref{fig:fullBlown} for expanded form of the equations).

The synchrotron frequency and power per electron are given by: 
\begin{equation}
	\nu_\text{sync} = \kappa_\nu \gamma^2 B,
\end{equation}
\begin{equation}
	P_{e} = \kappa_p \gamma_{\bot} ^2 B^2,
\end{equation}
where $\kappa_p=1.6 \times 10 ^{-15}\mbox{~ergs s$^{-1}$ G$^{-2}$}$ and 
$\kappa_\nu = 1.2\times 10^{6} \mbox{~Hz G$^{-1}$}$ are constants \citep{Lang}. 
For simplicity, we have used the synchrotron peak frequency in place 
of the sharply peaked complete distribution. 

If we assume the electrons only emit 
for a short time (either due to recombination or dissipative processes) then 
the number of synchrotron electrons is the same as the activity, $\mathscr{A}$. 
The activity is, of course, related to the decay constants of the unstable neutron-rich
r-process nuclei: 
\begin{equation}
	N_\textrm{total}(t) = \sum_i^\textit{chain} \frac{\mathscr{A}_i}{\lambda_i}.
\end{equation}
We write this explicitly to emphasize that there are many 
nuclear species present at the end of the r-process 
and there are many decay rates simultaneously contributing. 
This makes it difficult to constrain individual 
$\beta$-decay rates, but can help to constrain nuclear theory
models (not unlike the r-process).

\subsection{Summary of Model Components}
Our model involves many individual components as outlined above 
in sections~\ref{sec:firstPiece}-\ref{sec:lastPiece} and so 
we will shortly summarize how each qualitatively affects the 
emitted spectra. The r-process and fission recycling generate 
the electrons required for synchrotron 
emission and light cylinder provides a valve-like disruption and 
a monopole magnetic field. The spectrum is affected by the 
characteristics of the geometry and nuclear decay that exists 
in a QN. The exponential decay generates an exponential 
pulse at each particular emission frequency; whereas, 
the period of the neutron star affects magnetic field strength 
that the nuclei see as they cross the LC. In this simple model, 
the QN ejecta emits at decreasing energies and frequencies as the 
latitude of intersection increases and the exact bandwidth of the emission is 
determined by the parent neutron star's period and and LC field. 
In Fig.~\ref{fig:fullBlown} we show a total expression for the 
emitted spectrum.

\section{Total Power and Spectrum}
\label{sec:puttingItTogether}
The total number of nuclei and by extension electron emitters at a particular latitude is expressed as: 
\begin{equation}
	N_e = \sum_i\left( \frac{\mathscr{A}_i}{\lambda_i}\right) \int_{\phi_1}^{\phi} \cos \phi' d\phi' 
 = \sum_i\left( \frac{\mathscr{A}_i}{\lambda_i}\right) (\sin \phi - \sin \phi_1).
\end{equation}
The number of emitting electrons is then conveniently defined as a fraction of 
the total activity of the ejecta which is related to the total number 
of initial radioactive isotopes as defined in Eq.~\ref{eq:nucDecay}. This is simply the 
total number of isotopes in the quark nova ejecta and is given by:
\begin{equation}
	N_0 = \frac{m_{\rm QN,ejecta} }{ A_\textit{avg} m_H}.
\end{equation}
The power then decays exponentially in time according to the 
activity of the nuclear species in the ejecta as it 
passes through the LC. 

In order to determine the total power 
over the entire area we integrate the polar angle and latitude while 
keeping in mind that there is a time delay associated with the arrival of each 
latitude angle, $\phi$ due to the intersection of the spherical 
shock wave with the LC (giving us the Heaviside theta function, $\Theta$).

We can then construct an expression for the instantaneous power and spectral flux density as follows:
\begin{align}
	\mathscr{P}  &= \Theta(t) P_e N_e , \\
	\mathscr{S}  &= \frac{\Theta(t) P_e N_e}{\nu},
\end{align}
which are all evaluated at the arrival time (Eq.~\ref{eq:arrivalTime}) 
which has an angular dependence.
These expressions can then can be integrated to get the total 
power or spectral flux density: 
\begin{align}
	P(t) &= \oiint\limits_{S(\theta, \phi)} dP  = 
		\oiint\limits_{S(\theta, \phi)} 
		\left[ \frac{\partial \mathscr{P}}{\partial \phi} \right] d\Omega, \\
	S(t) &= \oiint\limits_{S(\theta, \phi)} dS = 
		\oiint\limits_{S(\theta, \phi)} 
		\left[ \frac{\partial \mathscr{S}}{\partial \phi} \right] d\Omega, 
\end{align}
which are both very unfriendly looking integrands; however, they both 
only have functional dependence on latitude angle, $\phi$. 

It is worth noting here that these 
integrals are evaluated at the source (in a frame near the LC). 
This means that cosmological redshift and dispersion are not accounted 
for in these expressions. While redshift is not included in this paper, 
new smaller dispersion measures are computed using the intrinsic time 
delay between emitted frequencies given by our model which 
traces out the magnetic field strength of the LC. These power and 
flux equations  
can of course both be integrated with respect to time to 
compute the total energy and the fluence respectively. 

\begin{figure}
	\centering
	\includegraphics[width=\linewidth]{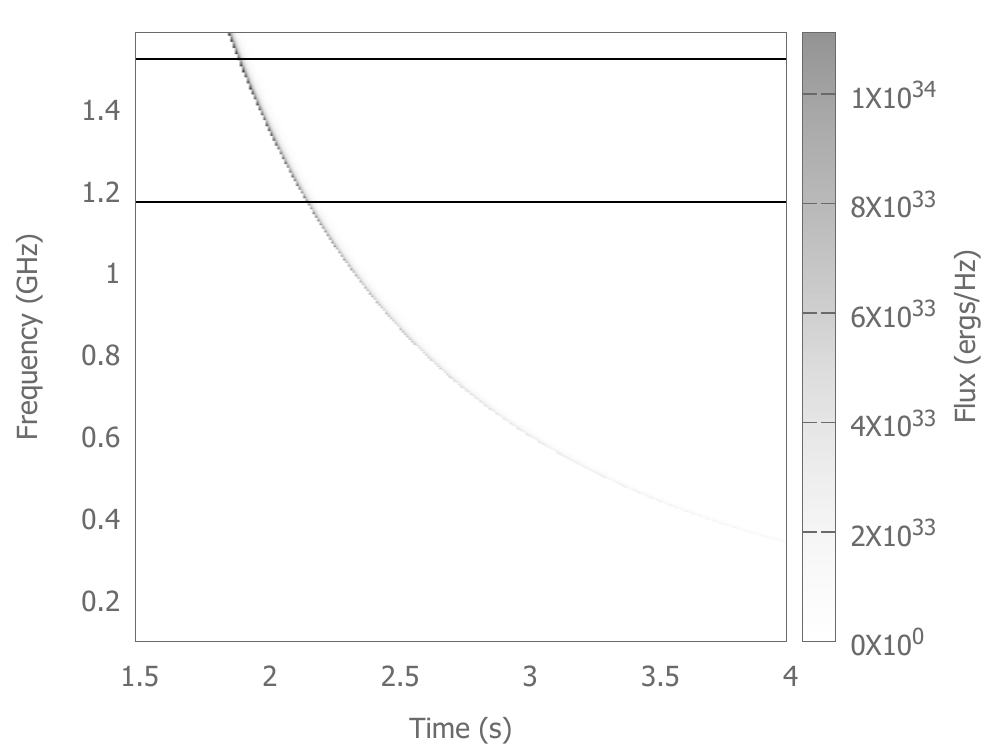}
	\caption{A spectrograph of the flux density at the emission location. The 
	upper edge of the spectrograph is a cut-off frequency determined by the 
	transition angle of the magnetic field where the magnetic field becomes mono-polar
	($\phi=30^\circ$). The two horizontal lines indicate the bandwidth 
	which has been measured by  \citep{Thornton2013}. This plot is 
	a fit to data using the 
	parameters outlined in Table~\ref{table:params} (no statistical 
	fitting was performed to find best 
	fits to the data).  \label{fig:sourceSpectrograph}}
\end{figure}

\begin{figure}
	\centering
	\includegraphics[width=\linewidth]{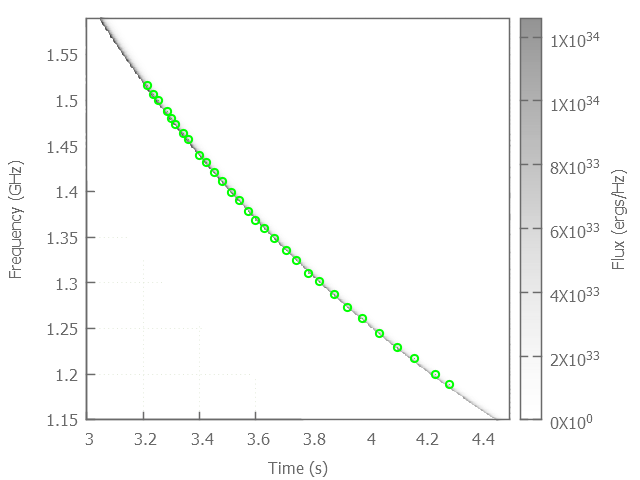}
	\caption{A spectrograph of the measured data (green points) 
	with our theoretical model predictions. A dispersion measure (DM) of 725~cm$^{-3}$~pc 
	has been applied to account for the time delay discrepancy between the 
	emission at source and measurement. The green circles 
	on the plot show where the observational data peaks 
	in the spectrograph.  \label{fig:dispersionedSpectrograph}}
\end{figure}

\begin{figure}
	\centering
	\includegraphics[trim=0 50 0 0,clip,width=\linewidth]{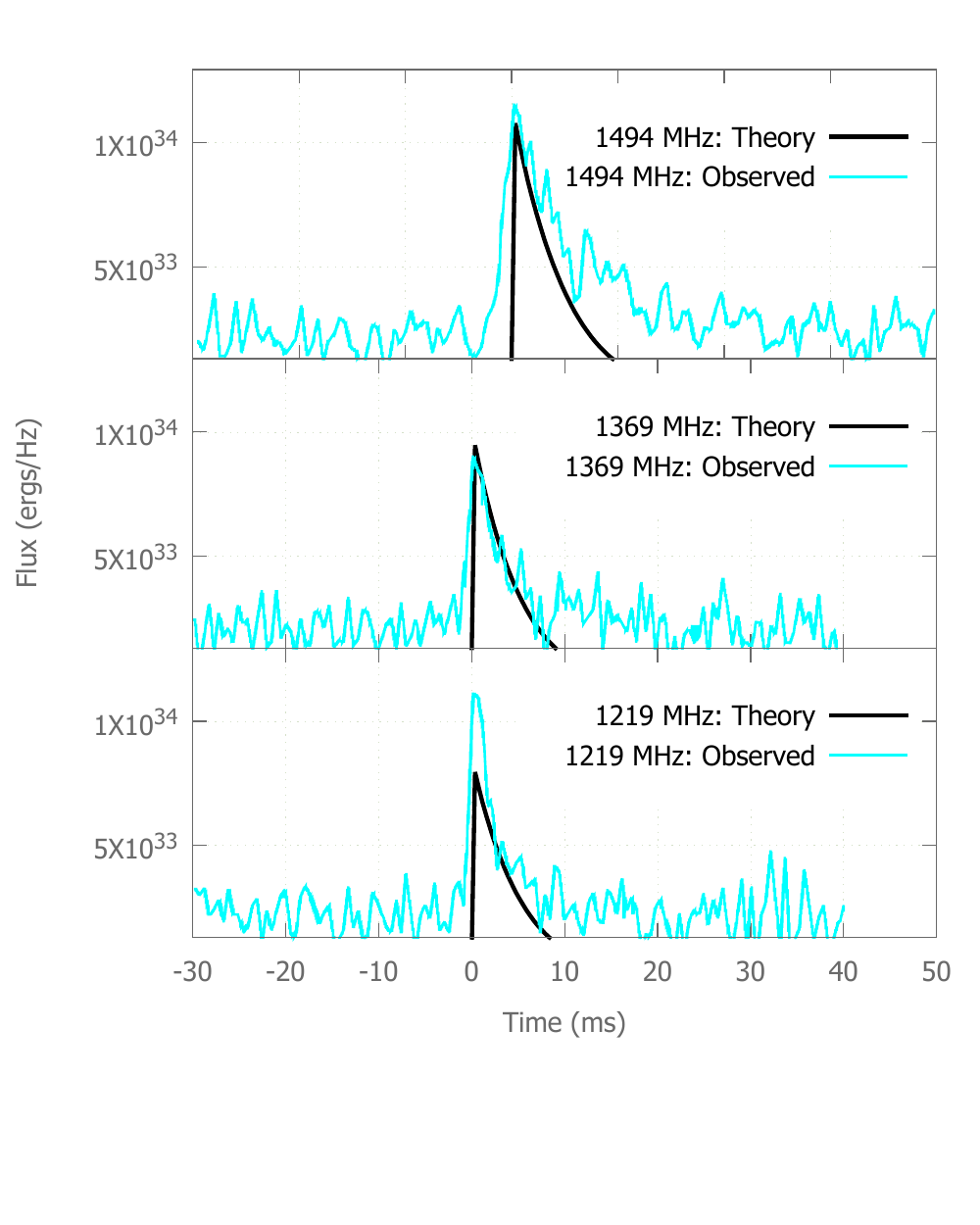}
	\caption{Slices into the spectrograph of Fig.~\ref{fig:sourceSpectrograph} at 1.494, 1.369 and 1.219 GHz (from 
	left to right). The relative height (or flatness) of the flux at these frequencies across the observing 
	band provides a constraint on the period of the neutron star progenitor. 
	This plot is 
	a fit to data using the 
	parameters outlined in Table~\ref{table:params} (no statistical 
	fitting was performed to find best 
	fits to the data).  \label{fig:slices}}
\end{figure}

\begin{figure*}
\begin{framed}
\begin{equation}
	P(t=t_{arr}, \theta, \phi ) = \oiint\limits_{S(\theta, \phi)}
		\frac{\partial}{\partial \phi}\left[
		\Theta(t) \kappa_p \gamma_{\bot}^2 B_\textit{NS}^2 
		\left( \frac{2\pi r_\textit{NS} \cos(\phi)}{c P }  \right)^{2a} 
		\sum_i^\textit{chain} \left( \frac{\mathscr{A}_n(t)}{\lambda_n}\right) 
		(\sin \phi - \sin \phi_1)
		\right] d\Omega
\end{equation}

\begin{equation}
	S(t=t_{arr}, \theta, \phi ) = \oiint\limits_{S(\theta, \phi)}
		\frac{\partial}{\partial \phi}\left[
		\Theta(t) \frac{\kappa_p }{\kappa_\nu} B_0
		\left( \frac{2\pi r_0 \cos(\phi)}{c P }  \right)^{a} 
		\sum_i^\textit{chain} \left( \frac{\mathscr{A}_n(t)}{\lambda_n}\right) 
		(\sin \phi - \sin \phi_1)
		\right] d\Omega
\end{equation}
\end{framed}
\caption{Expanded expressions detailing the functional form of the integrands describing 
the emission spectrum of the QN FRB model. The expression are left unevaluated 
as derivatives because they become lengthy and do not add any intuition to the problem. \label{fig:fullBlown}}
\end{figure*}

\begin{table}[h]
\centering
\begin{tabular}{ll}
\textbf{Parameter}  & \textbf{Value used}                          \\ \hline
$B_0$               & $10^{12}$ G                     \\
$\phi_{cut}$        & $30^\circ$                                   \\
P                   & 10 s                                         \\
\{$\lambda_i$\}     & \{1004, 855, 597, 425, 320\} s$^{-1}$ \\
$\beta$             & 0.99                                         \\
$\gamma$            & 2                                            \\
$r_0$               & 10 km                                        \\
$m_{\rm QN,ejecta}$ & $10^{-3} $M$_{\odot}$                          \\
$A_{\rm avg}$       & 180                                         
\end{tabular}
\caption{Summary of parameter values used in the Fig.~\ref{fig:sourceSpectrograph} and Fig.~\ref{fig:slices}. \label{table:params}} 
\end{table}

\subsection{Model Assumptions and Limitations}
The model has two major assumptions: 1) the high density ejecta from the QN 
maintains a distribution of unstable nuclei for far longer than is normally thought possible and 
2) the interaction at the LC causes a disruption of the r-process distribution. 
The r-process stationary distribution and its lifetime have been simulated using 
the nucleosynthesis code 
r-Java 2.0\footnote{r-Java 2.0 is an open 
use nucleosynthesis code developed at the University of Calgary at quarknova.ca \citep{rJava2} }. While it is a surprising result, we have theoretical 
background and simulation results to show that it is theoretically possible. The second 
assumption is much more difficult to quantify. To our knowledge, there has been no 
theoretical work done on the behaviour of dense nuclear material as it passes at 
relativistic velocities across a LC. This would require complex 
magneto-hydrodynamic modelling of material near what is already a difficult numerical 
problem. In order for the model to work, we require a disruption of the r-process distribution
at the LC. The most straightforward way to accomplish this 
is with rapid decompression of the material as shown 
quantitatively by the following equation: 
\begin{equation}
	r_{(n,\gamma)} = n_n \langle \sigma \rangle Y_i,
\end{equation}
where $r_{(n,\gamma)}$ is the neutron capture rate of isotope i, 
$n_n$ is the neutron number density, $ \langle \sigma \rangle $
is the average cross section and $Y_i$ is the number of isotopes in the system. 
If the ejecta is rapidly decompressed at the LC this 
would lead to a rapid decay of all isotopes to stable isotopes. 

Other approximations include: a symmetric spherical thin shell expansion of 
the ejecta, simplified electron energy distributions and use of 
theoretical decay rates, simplification of the 
synchrotron radiation spectrum of an electron, and a simplified 
selection for $\beta$-decay rates. 
These are each worth exploring in detail to provide a more accurate 
model; however, the agreement with observations leads us to conclude 
that these assumptions are all appropriate and physically valid. 

\section{Results}
Application of the model to the radio burst 
FRB~110220--the most spectroscopically complete FRB 
in \citet{Thornton2013}--is presented in Figs.~\ref{fig:sourceSpectrograph},  \ref{fig:dispersionedSpectrograph}, and \ref{fig:slices} using 
model parameters summarized in Table~\ref{table:params}. The total 
spectrum emitted using these parameters is presented 
in Fig.~\ref{fig:sourceSpectrograph}. The emission has a maximum 
frequency and decays for the lower frequencies over time. A comparison 
to the measured data of FRB~110220 (after a newly inferred 
dispersion measure is applied) is shown in 
Figs.~\ref{fig:dispersionedSpectrograph} and \ref{fig:slices}. 

\subsection{Dispersion Measure, Burst Profile and Distance}
The geometry and fields provide inherent cut-off ranges 
and characteristic emission timescales for our 
emitted spectra. 
The sudden turn-on (as represented by our Heaviside function) 
at the LC provides us with a frequency leading edge
which depends on magnetic field strength, period and - to a lesser extent - 
magnetic field configuration. The leading edge of the pulse for each frequency 
(as seen in a spectrograph) is a direct consequence of the geometry. The 
curve traced out in frequency over time is given by: 
\begin{equation}
	\nu_\mathrm{edge} = \kappa_\nu \gamma^2 B_0 \left( \frac{r_0}{\beta c} \right)^a \frac{1}{t^a},
\end{equation}
where parameter $a$ is the magnetic field configuration power. 
Using this, 
we get a time delay \emph{at the source} between two frequencies. This 
can be combined with observations to calculate a dispersion measure 
of the pulse which will be smaller than that calculated assuming 
synchronous emission. For FRB~110220 this leads to a DM of 725~cm$^{-3}$~pc as fit in Fig.~\ref{fig:dispersionedSpectrograph} 
which is reduction from 944~cm$^{-3}$~pc as measured in \citet{Thornton2013}.

There is an ambiguity in fitting a dispersion measure from the pulse 
time which is resolved by the magnetic field strength and the breakpoint 
between dipole and monopole field lines. Since the radiation emitted 
at the equatorial region is so much fainter, the maximum frequency of 
the measured spectra is given by:
\begin{equation}
\label{eq:maxFreq}
	\nu_{max} = \kappa_\nu \gamma^2 B_0 
		\left( \frac{2 \pi r_0 \cos\phi_{cut}}{c P} \right)^2.
\end{equation}
This can be used to help infer the correct dispersion measure of the 
measured radio burst. This, in combination with the spectral flux 
at each frequency can be used to calculate a more accurate dispersion measure 
and distance. The P$^{-2}$ dependence in Eq.~\ref{eq:maxFreq} leads to a 
wide range of possible frequencies. For example, given a millisecond period neutron star (P~$=1$~ms, B~$=10^{14}$~G) the peak 
frequency would be $16\times10^{18}$ Hz putting it in the X-ray regime. This extreme case is unlikely to be 
realized, as discussed in Section~\ref{sec:intro}. More realistic parameters suggest that FRB-like spectra are  
present in the optical and infrared bands, which depends on the period and magnetic 
field strength of FRB progenitors. Determination of the likelihood of emission at 
different frequencies from this mechanism 
would require detailed neutron star population studies aimed specifically at determining FRB-QN rates.

In tandem with the new dispersion measure, we can estimate the 
distance to these QNe-FRBs by integrating the flux density 
or power spectra and comparing this to the measured energy or 
fluence. This is possible even for bandwidth limited measurements 
provided there is an estimate from the above quantities of the period 
and magnetic field by integrating only over the same range of frequencies 
present in the measurement. If these bursts are in fact cosmological, 
the cosmological redshift will have to be incorporated into the above 
equations.

\subsection{FRB Rates}
\label{sec:rates}
Assuming that FRB progenitors are old neutron stars, we can estimate their rate based on a
simplified population model. We give a rough estimate of the FRB rate
by making use of the following assumptions: a constant core-collapse supernova rate 
(0.02 yr$^{-1}$ galaxy$^{-1}$), a lognormal magnetic field distribution 
($\chi(\log\textrm{B}) = 13.43\pm0.76$) and a normal distribution for the initial period 
($\pi(\textrm{P}_0) = 290\pm100$ ms), where the values for the initial period and magnetic field 
are taken from \citet{2001A&A...374..182R} (the magnetic field distribution is from the ``unseen'' 
population of Model A). In order to decide the lifetime of the neutron 
stars when they go FRB we've made two choices: 1) they go QN-FRB after 
a common timescale (or age), or 2) they go QN-FRB once they reach 
a certain location in the P$\dot{\textrm{P}}$ diagram. For the 
second case, we consider neutron stars reaching critical 
density near the observed death line and employ the death line of \citet{2000ApJ...531L.135Z}:
\begin{equation}
\log\left( \frac{\dot{P}}{\textrm{s s}^{-1}} \right) = 2\log\left( \frac{P}{\textrm{s}} \right) -16.52.
\end{equation}
After modelling the neutron star population of a Milky-Way-like galaxy, 
the last step is to estimate the fraction of neutron stars which are QN candidates. 
Since only slow rotating, massive neutron stars are FRB candidates, we count only 
the neutron stars with periods greater than 100 ms and whose stellar progenitors had 
masses between 20--40 M$_{\sun}$. Integration of the 
IMF \citep{2001MNRAS.322..231K} and the initial period distribution (assuming the period  and 
magnetic field are independent) gives approximately 14\% of neutron stars as QN candidates. 
This approach does not take into account any possible viewing selection effects (e.g. threshold observing brightness, viewing angle, 
beaming, etc.), which may affect the detectability of these events. Both the period and death line estimates of the 
rate give consistent results of approximately 3 per thousand year per galaxy, as summarized in 
Table~\ref{table:rates}, which is consistent with the observed FRB rate.  

\begin{table}[h]
\centering
\begin{tabular}{ll}
\bf{Critical Period (s)} & \bf{Rate (yr$^{-1}$ galaxy$^{-1})$} \\ \hline
1 & 0.0028 \\
10 & 0.0027 \\
Death Line & 0.0028
\end{tabular}
\caption{ Summary of FRB rates calculated using a simplistic population synthesis of 
neutron star population. \label{table:rates}}
\end{table}

\section{Model consequences and predictions} 
While we can generate excellent fits to currently available FRB 
data, there are other important observables and consequences 
that come along with the scenario we have put forth to explain 
these radio transients. As mentioned earlier, we can use this 
model to constrain certain physical and astronomical phenomena. 
The relation to nuclear properties is somewhat complex due to 
the number of nuclear species present and is best left 
for more detailed analysis and higher resolution measurements. There are, however, 
two astronomical puzzles that we believe can be explained 
assuming this model is in fact correct.

\subsection{Neutron Star Population}
Assuming the model is correct, FRBs provide 
measurements of the period and magnetic field strength of isolated 
NS before they undergo a QN. Based on the agreement of the 
model with FRB~110220 and the spectral consistency of the FRB measurements, 
it appears that these are neutron stars that undergo QN.
These neutron stars have periods of around 10 
seconds and magnetic fields of $10^{12}$~Gauss (as determined by 
our parameter selection and its fit to observation data). 
This is perhaps puzzling 
if we look at the so-called P$\dot{\textrm{P}}$ diagram: there are no 
observed neutron stars in this 
region \citep{PPdotBook}. This empty region of the 
P$\dot{\textrm{P}}$ diagram is beyond what is referred to as the ``death line'' 
and we propose an explanation for this: once an isolated neutron 
star spins down to the 1 to 10 second period and enters the neutron 
star graveyard, it reaches 
deconfinement densities at its core and dies via a QN explosion. 
This instantaneously moves the neutron star into the $10^{15}$~Gauss 
regime of the diagram as quark stars generate these high magnetic 
fields after birth \citep{Iwazaki}. Instead of magnetars, we then populate this 
anomalous region of the P$\dot{\textrm{P}}$ diagram with quark stars whose 
magnetic fields decay not through spin down, but through 
magnetic field expulsion from the color superconducting core \citep{NiebergalVortexExpulsion,QSBPCXRay}. This 
drives the quark star from its new position straight downwards as 
the star ages and provides an explanation for the observed high magnetic field 
population in the P$\dot{\textrm{P}}$ diagram.

\subsection{511 keV Line and Super-heavy Fissionable Elements}
The production of super-heavy ($A>240, Z>92$) elements is responsible 
for the fission recycling in the QN ejecta. While these calculations 
are based on theoretical predictions of mass models, fission recycling from 
these elements is theoretically expected. Some of these super-heavy nuclei 
are also thought to be quasi-stable and long-lived. While they are (currently) 
impossible to make in the lab, these neutron-rich super-heavy nuclei may 
outlive the $\beta$-decaying nuclei and undergo fission much later. 
If this is correct, these super-heavy nuclei would fission and emit 
photons over a long period of time and produce extended emission of 
MeV photons. These emitted photons then have the appropriate energy 
for efficient production of electron-positron pairs 
(via $\gamma$-$\gamma$ 
annihilation). The annihilation events would produce 511~keV photons 
when the produced positrons find an electron and 
provide a potential explanation for the positron emission which 
is an observed phenomenon that still has no definitive explanation 
for its galactic origin \citep{Weidenspointner2008}.

This radiation would of course not be as energetic as the original FRB and 
is likely not measurable from an extragalactic source; however, this does 
provide a  possible explanation for the 511~keV photons observed in our own galaxy. 
The current observed annihilation rate of positrons inferred from the 
511 keV positronium line is $\sim 10^{43}$ s$^{-1}$ \citep{KnodlsederAllSkyPositron}. 
Based on our model, 
we can then constrain the lifetime of super-heavy neutron rich nuclei 
to be on the order of 1000 years using the following logic.

Given a galactic QN rate of 1 per thousand years, if we assume the 
following parameters: $M_{\rm QN,ejected}$ = $10^{-2}$ M$_\odot$, 
average nuclei mass A$\sim 100$ and 10\% of nuclei 
(by number) are super-heavy fissionable elements, 
we can expect about $10^{-5}$ $M_\odot$ nuclei produced in the QN ejecta 
which gives about $10^{51}$ fissionable nuclei per QN event. Given the 
expected QN rate of 0.001 year$^{-1}$ this produces  an 
average $10^{48}$ super-heavy nuclei per year. If we compare this to the current event rate this implies 
a characteristic lifetime of about 1000 years for the fission events 
(see \citet{Petroff2015} for further discussion of observational qualities and
proposed astrophysical theories). 

While these calculations are an order of magnitude approximation, it has 
several promising features. QNe come from massive (20-40 $M_\odot$) stars and are expected 
to operate (like the r-process) continually over the entire history of the galaxy. 
This leads naturally to a concentration of 511 keV lines near the active portions 
of the galaxy (the bulge and the disk) where most of this emission is observed. This 
matches the observed clustering of the 511 positron emission measured \citep{Weidenspointner2008}.
Additionally, because a quark star is an aligned rotator, we expect no radio pulsations to be
observed alongside this emission which, again, seems to agree with observation and is one 
of the difficult aspects in explaining these positronium annihilation events. 
Finally, 
since the galactic centre and plane has many high energy photons which would allow 
for efficient conversion of the MeV photons from fission to electron-positron pairs 
(because the pair production requires a $\gamma$-$\gamma$ interaction).
This emission should then carry a P Cygni profile as well as a net 
exponential decay which could be detected from observations. 

It is also possible that the heavy nuclei do not spontaneously fission, 
but instead undergo photo-fission triggered by the high energy photons 
present in large numbers especially near the galactic centre. This would lead 
to a slightly different observable: instead of half-lives we could infer photo-fission
cross sections which would give indirect measurements of the level density and fission 
barrier heights of these nuclei. Instead of a plain exponential, we would instead observe 
rates proportional to the photon density. 

\section{Conclusion}
Quark novae are capable of generating fast radio pulses using 
electrons generated by r-process nuclei. Assuming decompression or any other 
disruption of the r-process 
at the neutron star's LC, we have shown how the emission 
spectrum has an intrinsic time delay across its frequency band and has 
a predictable bandwidth (cut-off frequency) which can be used to
measure magnetic field strength and period of otherwise undetectable  
neutron stars. If correct, FRBs can be used to probe and measure 
properties of unstable nuclear matter and measure properties 
of old neutron star populations. Our model predicts that these explosions are 
indeed one of the most energetic explosions in radio astronomy (with 
energies on the order of $10^{41}$~ergs); however, 
they need not necessarily be cosmological 
as is inferred from 
previous analysis of the data. The model naturally 
provides explanations for the timescales and energies of the FRB 
emissions using synchrotron emission in the strong magnetic field of 
a birthing quark nova.

\bibliographystyle{raa}
\bibliography{FRBbib}

\end{document}